\documentclass[pre,aps,floatfix]{revtex4}

\usepackage{amsmath,amssymb}
\usepackage{graphicx}
\usepackage{psfrag}
\usepackage{graphicx}

\def\beq{\begin{equation}}
\def\eeq{\end{equation}}
\def\bea{\begin{eqnarray}}
\def\eea{\end{eqnarray}}

\begin{document}

\title{
Generic instabilities in a fluid membrane coupled to a thin layer of
ordered active polar fluid}
\author{Niladri Sarkar and Abhik Basu} \affiliation{{Theoretical Condensed Matter
Physics Division, Saha Institute of Nuclear Physics, 1/AF,
Bidhannagar, Calcutta 700 064, India} }

\date{\today}

\begin{abstract}
We develop an effective two-dimensional coarse-grained description
for the coupled system of a planar fluid membrane anchored to a thin
layer of polar ordered active fluid below. The macroscopic
orientation of the active fluid layer is assumed to be perpendicular
to the attached membrane. We demonstrate that {\em activity} or
nonequilibrium drive of the active fluid makes such a system
generically linearly unstable for either signature of a model
parameter $\Delta\mu$ that characterises the strength of activity.
Depending upon boundary conditions and within a range of the
model parameters, underdamped propagating waves may be present in
our model. We discuss the phenomenological significance of our
results.
\end{abstract}

\maketitle

\section{Introduction}

Studies on fluid membranes have a long tradition in equilibrium
statistical mechanics~\cite{fluidmem-basic}. Coarse-grained
descriptions of a fluid membrane, parametrised by its tension,
bending modulus and spontaneous curvature~\cite{helfrich}, provide a
quantitatively accurate description of thermal fluctuations of fluid
membranes.    Plasma membranes in eukaryotic cells, ignoring their
bilayer structure and internal complications, are broadly described
by a fluid membrane, typically attached to a layer of cortical actin
filaments. Understanding of the mechanical responses of {\em
in-vivo} eukaryotic cells to external stimuli, in which both the
cell membrane and actin filaments are expected to play crucial
roles, remains a highly challenging subject due to the underlying
structural complexities of a cell. It is, therefore, useful to
theoretically consider or experimentally design simple {\em
in-vitro} systems involving membranes and actin filaments, whose
macroscopic physical properties may be understood or analysed in a
straightforward manner. In the spirit of this minimalist approach,
in this article we construct a coarse-grained two-dimensional ($2d$)
hydrodynamic description of the coupled dynamics of a fluid membrane
and a macroscopically polar ordered active fluid layer anchored
locally normally to it. In general, a combined system of a plasma
membrane and an anchored layer of actin filaments executes dynamics
that is distinctly nonequilibrium~\cite{noneq1,noneq2}. Hence,
thermal equilibrium models of fluid membrane do not suffice as
meaningful descriptions of such a system.

There have been several recent studies on the various aspects of out
of equilibrium dynamics of fluid membranes in systems with
biological relevance; see, e.g.,
Refs.~\cite{sriram-pump,nir,nir1,zimm,sumithra,niladri0,niladri,bead}.
 In the present work, we complement the existing results and
consider a thin layer of active fluid with the active particles
normally anchored to a fluid membrane at length scales much larger
than the lengths of the individual polar particles (e.g., actin
filaments) and layer thickness with polar ordering, for which a
generic coarse-grained continuum two-dimensional ($2d$) description
should suffice.   The nonequilibrium aspect of the active fluid
dynamics is modelled in terms of an active contribution to the stress
tensor, called {\em active stress} (nonequilibrium stress), proposed
and elucidated in Refs.~\cite{aditi,kruse}; see also
Ref.~\cite{reviews} for recent development in the subject. It is
assumed that the active fluid layer is covered by a fluid membrane
on one side, which is parametrised by its surface tension and
bending stiffness only. Here, we
   systematically construct a set of thickness-averaged
effective $2d$ coarse-grained coupled dynamical equations for our
model, by following the approaches outlined in
Refs.~\cite{sumithra,niladri}. We use our equations to study linear
instabilities about the assumed polar ordered state. We find that
the system  shows generic instability for both signatures of
$\Delta\mu$, a parameter that characterises the strength of the
active stress in the model. In addition, under certain circumstances
and depending upon the boundary conditions imposed, the system may
display underdamped propagating waves as well. While our
calculational framework is closely related to Ref.~\cite{niladri}
(see also Ref.~\cite{sumithra}), the system under consideration here
is quite different from those in Refs.~\cite{sumithra,niladri} in
having a macroscopic orientation different from
Refs.~\cite{sumithra,niladri}. Furthermore, we include in our model
a second active species, represented by a local active density, that
exists on the membrane. This may, for instance, model
 proteins that bind with the actin filaments and facilitate actin
 polymerisation.

 Our model, although lacks many biological details, is motivated by the consideration that the
cortical actin layer is generally isotropic in-plane, and is
anchored to the cell membrane. Technically, if the underlying polar
particles are actin filaments, our description should be valid for
time scales larger than the unbinding time scale of the
cross-linking proteins of the actin filaments, so that the actin
network behaves like a fluid.  Furthermore, our model should be
useful as a starting point for theoretical descriptions for recent
{\em in-vitro} experiments~\cite{vitro,bass,koend}. Despite the
technical difficulties involved, our results described here may in
principle be tested by performing controlled experiments on thin
confined actin layers, grafted normally on both the bounding
surfaces with a sample thickness much smaller than the correlation
length in the thin direction. The rest of the paper is organised as
follow: In Sec.~\ref{model}, we set up our model, specify the
boundary conditions and derive the $2d$ equations of motion. In
Sec.~\ref{instabil} we perform linear stability analysis on our
model equations. In Sec.~\ref{conclu} we conclude and discuss our
results.

\section{Construction of the model and the equations of motion}
\label{model}

Assuming a planar membrane for simplicity, we consider a thin layer
of ordered active fluid of viscosity $\eta$, attached to the
membrane, spread in the $xy$ plane and thin in the $z$-direction.
Our study begins with the three-dimensional ($3d$) hydrodynamic
description of a polar ordered state in this model system; the
relevant dynamical fields include (we assume an incompressible
system) local velocity field ${\bf v}({\bf r})= (v_x,v_y,v_z)$, a
polarisation field ${\bf p}({\bf r})=(p_x,p_y,p_z)$, a unit vector
that describes any local orientational order, concentration
$c(x,y,z)$ of the active particles and $\psi(x,y)$ which describes
an active species density that exists on the membrane. Our system is
confined between the surfaces $z=h_1$ and $z=h_2$. We study small
fluctuations about a chosen reference state given by $p_z=1$.
Our aim is to construct an {\em effective} $2d$ description of the
combined active polar fluid-membrane dynamics, such that any
$z$-dependence is averaged out and the effective dynamical variables
depend only on $x,y$. Similar to Ref.~\cite{niladri}, we consider
two different versions of our model: (i) Model I: The fluid
membrane-active fluid combine rests on a solid substrate below, and
(ii) Model II: The system is embedded inside a bulk isotropic
passive fluid having a viscosity $\eta' $, assumed to be much
smaller than $\eta$. As discussed in details in Ref.~\cite{niladri},
the two cases are physically different. The solid substrate below
introduces friction and, hence, the momentum (or the hydrodynamic
velocity $\bf v$ for an incompressible system) is no longer a
conserved variable. In contrast, with a bulk fluid surrounding the
system there is no friction at the interface, and therefore, the
momentum density (equivalently, $\bf v$ for an incompressible
system) is a conserved variable. Furthermore, the solid substrate
breaks the full three-dimensional ($3d$) rotational invariance of
the problem, where as, when there is an embedding bulk fluid, the
system remains invariant under the full $3d$ rotational invariance
(see below). We separately discuss Model I and Model II in details
below, and construct coupled equations motion for the membrane
height field, velocity field of the active fluid, polar particle
concentration and the active density on the membrane.


\subsection{Model I: The membrane active gel combine bounded by a
solid substrate below}

\subsubsection{Construction of the model and the free energy
functional}

\begin{figure}[htb]
 \includegraphics[height=6cm]{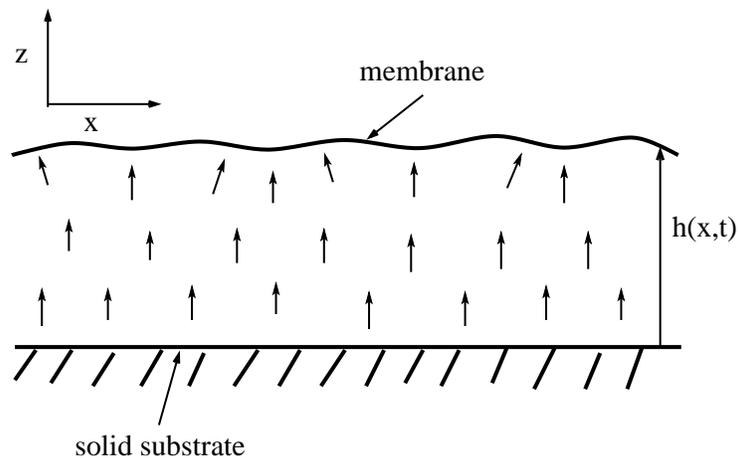}
 \caption{A schematic diagram of our model system showing the
membrane and the active fluid layer resting on a solid substrate.
The arrows indicate the direction of macroscopic orientation (here
along the $z$-axis).}
 \label{solfig}
 \end{figure}

 Model I,  where the fluid membrane-active fluid layer
combine rests on a fully flat solid substrate (see
Fig.~\ref{solfig}), is relevant in the context of {\em in-vitro}
experiments discussed, e.g., in Ref.~\cite{vitro}.
 In order to set  up the equations of motion we
first consider the form of the relevant free energy functional
${\mathcal F}_s$ that gives the energy of a configuration defined by
height field $h_1=h$ (we set $h_2=0$, since it is a rigid surface
below) and polarisation $\bf p$. The form of ${\mathcal F}_s$ may be
inferred by using symmetry arguments. First of all, the presence of
the solid substrate below makes the system invariant under $2d$
in-plane ($XY$-plane) rotation and the translation $h\rightarrow h +
const.$ (we neglect any interaction between the membrane and the
bottom solid surface, which is reasonable if the average thickness
of the intervening active fluid is not too small). However, there is
no invariance under a full $3d$ rotation. As discussed in
Ref.~\cite{niladri,aditi}, the symmetry considerations as above
dictates that the leading order (in gradients) coupling between $\bf
p$ and $h$ and the height of the membrane should be a term of the
form $\sim \int {\bf p}_\perp\cdot{\boldsymbol\nabla}_\bot h$, where
${\bf p}_\perp = (p_x,p_y)$. The chosen form of the ${\bf p}-h$
coupling is determined by the invariance under $2d$ in-plane
rotation and translation along $h$. This is a polar term, since it
has no $\bf p \rightarrow -p$ symmetry. Further, it violates the
$\bf h \rightarrow - h$ symmetry as well, which is admissible since
the active fluid is anchored only on one side of the membrane.
Further, the configuration energy of the membrane is assumed to be
determined by its local curvature and surface tension. Lastly, the
free energy of the active polar particles is given by the Frank free
energy for nematic liquid crystals~\cite{jacques-book}. Thus we have
 \bea
\mathcal{F}_s &=& {1 \over 2}\int d^2x\left[\kappa
(\nabla_\bot^2h)^2 + \sigma ({\boldsymbol\nabla}_\bot h)^2 + A(\psi
-\psi_0)^2- 2\lambda\psi\nabla_\bot^2h\right]
\nonumber \\
&& +{1 \over 2}\int d^2r\int_0^h dz [-\lambda_1 {\bf
p}_\perp\cdot{\boldsymbol\nabla}_\bot h \delta(z-h)- 2\lambda_2 {\bf
p}_\perp\cdot{\boldsymbol\nabla}_\bot \psi\delta(z-h) +
D(\nabla_ip_j)^2]. \label{freesol}\eea
 where  $\sigma$ and $\kappa$ are the surface
tension and bending modulus of the membrane, $\psi_0$ is the spatial
average of $\psi$, $A\sim T\psi_0$ is the osmotic modulus of the
membrane bound active density for small $\psi_0$, $T$ is the
temperature, $\lambda,\lambda_1$ and $\lambda_2$ are coupling
constants (all chosen to be positive) for the bilinear terms. We
impose $p^2=1$. The Frank free energy is considered in the limit of
equal Frank's constants, denoted by $D$ above. We have used the
Monge gauge \cite{monge} for the membrane height field in the free
energy functional (\ref{freesol}) and have kept only the terms which
are either quadratic or bilinear in the fields. Operator
${\boldsymbol \nabla}=(\frac{\partial}{\partial
x},\frac{\partial}{\partial y},\frac{\partial}{\partial z})$ and
${\boldsymbol\nabla}_\perp= (\frac{\partial}{\partial
x},\frac{\partial}{\partial y})$ are the $3d$ and $2d$ Laplacians
respectively.  All the parameters in the model are chosen in such a
way, that a stable spatially uniform equilibrium phase~\cite{equl}
(at zero activity) is ensured, e.g., $\lambda_1,\,\lambda_2$ are
assumed to be positive. Note that the free energy functional
(\ref{freesol}), with $\psi=0$, appears similar to the corresponding
free energy functional used in Ref.~\cite{niladri}. This is
essentially due to the polar nature of the systems considered in
Ref.~\cite{niladri} as well as here. A careful consideration,
however, reveals differences. For instance, the free energy and the
corresponding macroscopic behaviour in Ref.~\cite{niladri} have $2d$
in-plane anisotropy (although anisotropic coefficients were not used
in the free energy functional there, in order to reduce the
algebraic complications), due to the chosen ordered state with
$p_x=1$, where as the present model is invariant under $2d$ rotation
in the $XY$-plane. Due to the macroscopic orientation along the
$z$-direction, our system is, however, generally anisotropic in the
$z$-direction, resulting into anisotropic coupling constants, e.g.,
Frank's elastic constants. For reasons of simplicity and analytical
tractability, we choose to ignore such complications. Higher order
non-linearities should also reflect this anisotropy; since we
restrict ourselves only up to terms bilinear in the fields, we are
not concerned by such issues here.

\subsubsection{Boundary conditions}

Having constructed the free energy ${\mathcal F}_s$ above, we now
consider the boundary conditions to be imposed: (i) First of all,
the solid substrate below introduces friction; consequently, the
relevant boundary condition on the velocity field at $z=0$ is the
no-slip boundary condition on ${\bf v}_\perp$: ${\bf v}_\perp =0$,
where ${\bf v}_\perp = (v_x,v_y)$ and no penetration on $v_z$:
$v_z=0$ at $z=0$, (ii) Further, local orientation $\bf p$ is
constrained to be normal to the plane $z=0$ and the local normal at
$z=h$, and finally (iii) vanishing of the shear stress and the
normal stress balance at $z=h$. Assuming that the relevant
nonequilibrium intrinsic stress field of the active
particles~\cite{reviews} is
$\sigma_{\alpha\beta}^a,\alpha,\beta=x,y,z$, the vanishing of the
shear stress at $z=h$ yields
\begin{equation}
\eta(\partial_z v_i +\partial_i v_z)_{z=h}=\sigma^a_{iz}, i=x,y
\end{equation}
which, in the thin film approximation where $v_z$ is assumed to be
small and may be neglected, reduces to
\begin{equation}
\eta\left(\frac{\partial v_i}{\partial z}\right)_{z=h}=\sigma^a_{iz},\label{str}
\end{equation}
 at $z=h$, where $i=x,y$. The kinematic boundary condition together
with the incompressibility of the active fluid connects $h$ with the
flow:
\begin{equation}
{\partial h \over \partial t}=v_z=-\int
{\boldsymbol\nabla}_\bot\cdot{\bf v}_\bot dz. \label{kine}
\end{equation}
Considering active particles being normally grafted onto the
membrane, we have
 ${\bf p}\cdot{\hat N}=1$. This yields, for small fluctuations of the membrane, ${\bf p}_\bot=-2{\boldsymbol\nabla}_\bot h$ at $z=h$.
  Here $\hat N$ is the unit normal to the membrane surface, which in
the Monge gauge  is given by $\hat N\simeq
(-{\boldsymbol\nabla}_\perp h,1)$ to the lowest order in height
fluctuations. Since the surface at $z=0$ is  flat (a rigid solid
surface), the boundary condition on ${\bf p}_\bot$ at $z=0$ is ${\bf
p}_\bot =0$.

\subsubsection{Active stresses and the dynamical equations of motion}

The stress field $\sigma_{\alpha\beta}^a$ of the active
particles~\cite{reviews}, called {\em active stress} below, is of
the form
\begin{equation}
\sigma_{\alpha\beta}^a=\xi\Delta\mu c({\bf r})p_\alpha ({\bf
r})p_\beta({\bf r}) +\tilde\lambda \Delta\mu [\psi({\bf
r})-\psi_0]\delta (h-z)\hat {\bf z}\hat {\bf z}.\label{active}
\end{equation}
 In the context of eukaryotic cells, the constant parameter $\Delta\mu$,
 gives a measure of the free energy available from the hydrolysis
of the Adenosine Triphosphate (ATP) molecules inside the cell. For
bulk polar active fluids, with $\tilde\lambda =0$,
$\sigma_{\alpha\beta}^a$ is said to be contractile or extensile for
$\Delta\mu <0$ or $\Delta\mu
>0$~\cite{kruse}; its
numerical value characterises the strength of the active stress
field.  A local imbalance of the membrane-bound density $\psi$
creates a local normal stress component that is assumed to be
active. In (\ref{active}) we have ignored any coupling between
$\psi$ and the local curvature $\nabla^2 h$. Thus there are two
different sources for active stresses, the orientation field and the
local density $\psi$. The magnitudes of the couplings $\xi$ and
$\tilde\lambda$ (both assumed to be positive here) describe the
relative strengths of the different contributions to the active
stress (\ref{active}). In addition, we assume $c$ not to have any
significant $z$-dependence. Further, since we are interested in the
effective long wavelength dynamics in the $xy$-plane, in what
follows below, we use a linear profile for ${\bf
p}_\perp=-(2z{\boldsymbol \nabla}_\perp h)/h$ that clearly satisfies
the boundary conditions imposed on ${\bf p}_\perp$ and use this form
to calculate $z$-averaged active stresses~\cite{sumithra,zaverage}
from (\ref{active}): In particular we have for the shear active
stress
\begin{equation}
\langle\sigma_{iz}^a\rangle_z= -\xi\Delta\mu c_0 \partial_i h,
\label{zavgstr}
\end{equation}
where $\langle..\rangle_z$ implies $z$-averaging of any quantity,
$c_0 = \langle c\rangle$.
 This
allows us to borrow the methods of Refs.~\cite{niladri,sumithra}
directly for the present problem. How good is it expected to be? It
is well-known~\cite{bead,rafael}, at a critical thickness $\overline
h$, a Frederik-like spontaneous flow instability sets in. In the
model of Ref.~\cite{bead}, both the confining surfaces are held
fixed (non-fluctuating) and  ${\bf p}_\perp=0$ at both the top and
bottom boundaries, such that the profiles of ${\bf p}_\bot$ may be
written as a sum of appropriate trigonometric functions of $z$ that
obey ${\bf p}_\perp=0$ at $z=0,h$ automatically. With this, the
instability at $\overline h$, that depends upon $\Delta\mu$,
manifests itself explicitly. If the top surface becomes a flexible
surface with small fluctuations, we expect a small departure from
the $z$-dependent profiles for ${\bf p}_\bot$  in Ref.~\cite{bead}.
Therefore, our choice of a linear profile should not be a good
approximation for $h$ near $\overline h$; we expect our results here
will be meaningful in the limit $h\ll \overline h$ such that the
Frederik-like spontaneous flow instabilities are strongly suppressed
and consequently, a $z$-averaged description is physically valid.
Before embarking on our calculations, let us compare with
Ref.~\cite{niladri} briefly, where fluctuations about a state
$p_x=1$ is studied. There, $p_z$ is constrained to have prescribed
values at the boundaries (0 and $\sim\partial_x h$, at $z=0$ and
$z=h$ respectively), where as $p_y$ remains unconstrained at the
boundaries. As a result, $p_z$ is slaved to $h$ and drops out of the
effective $2d$ description that was constructed. In the present
model, for the same reason, $p_x$ and $p_y$, having specified values
at $z=h,0$, are slaved to $h$ and drop out of the eventual effective
$2d$ theory (moreover, to the lowest order in smallness, $p_z=1$
and, hence has no time evolution). Thus we are left out with
$h,\,\psi$ and $c$ as the slow variables that describe the dynamics
of the model in the long time limit.

As we shall see below, a full solution of the dynamics of the model
at hand entails solving for the coupled dynamics of $h,{\bf p},\psi$
and $c$ (this enters through the conservation of the active
particles, see below). Neglecting inertia, which is a good
approximation for active particles with small masses (equivalently,
low Reynolds number flows) and within the lubrication approximation,
the $3d$ velocity field $\bf v$ satisfies the generalised Stokes
equation that now includes the active stress term. In particular,
the in-plane component ${\bf v}_\perp$ obeys
 \bea \eta\frac{\partial ^2v_i}{\partial z^2} -
{\nabla}_i \Pi - {\nabla}_\beta \sigma_{i\beta}^a=0 \label{stokesS}
\eea where $\beta=x,y,z$, $i=x,y$.
 For the $z$-component of Eq.~(\ref{stokesS}) we use the
lubrication approximation for $v_z$ (i.e, $v_z\approx 0$) in the
thin film limit, which then yields  \bea
\partial_z\Pi=-{\nabla}_i\sigma^a_{iz}=\xi c_0\Delta\mu\nabla_\bot^2h,\label{pressS}
\eea
 where we have linearised about $p_z=1$ and
 $\psi_0 =\langle\psi\rangle$.
 Equation (\ref{pressS}) can be integrated over $z$ to solve for
the pressure $\Pi$. The constant of integration is to be fixed by
the condition of normal stress balance. The pressure at the location
of the membrane is balanced by the total normal stress at $z=h$.
This includes the elastic force of the membrane and the active
stress due to the active density.  Thus, $\Pi$ is obtained as
 \bea
\Pi= P_0 - \xi c_0\Delta\mu(h-z)\nabla_\bot^2h - f_s(h) +
\tilde\lambda\Delta\mu(\psi -\psi_0), \label{pisol} \eea where $f_s(h)$
is the elastic contribution to the stress which can be derived from
the free energy as $f_s(h)=-\delta\mathcal{F}_s/\delta h=-\kappa
\nabla_\bot^4h + \sigma\nabla_\bot^2h + \lambda_1 \nabla_\bot ^2h +
(\lambda+\lambda_2)\nabla_\bot^2\psi$, where we have replaced ${\bf
p}_\perp$ by its $z$-averaged form. Substituting for $\Pi$ from the
above in the linearised Stokes Eq. (\ref{stokesS}) for ${\bf
v}_\perp$, integrating twice with respect to $z$ and using boundary
conditions (i) at $z=0$ $v_i=0$ and (ii) at $z=h$, $\eta
\partial_z v_i = -\xi\Delta\mu c_0 \partial_i h$, we obtain
\begin{eqnarray}
\eta v_i &=& {\xi c_0\Delta\mu \over 2}\nabla_i (\frac{z^3}{3} - h^2
z)\nabla^2 h -\xi c_0\Delta\mu \nabla_i (\frac{z^2}{2} -
hz)h\nabla_\perp^2 h - (\lambda+\lambda_2)(\frac{z^2}{2}
-hz)\nabla_i \nabla_\bot^2 \psi \nonumber \\ &-&\xi c_0\Delta\mu
\frac{z^2}{h}\nabla_i h + \xi c_0\Delta\mu z\nabla_i h - \sigma
(\frac{z^2}{2} -hz) \nabla_i \nabla^2 h +\kappa
(\frac{z^2}{2}-hz)\nabla_i\nabla_\bot^4 h \nonumber \\ &+&
\tilde\lambda \Delta \mu (\frac{z^2}{2} - hz)\nabla_i\psi -
\lambda_1 \nabla_i \nabla^2 h (\frac{z^2}{2} -hz). \label{vperpsol}
\end{eqnarray}
We now use incompressibility of the fluid, giving $v_z=-\int_0^h dz
{\boldsymbol\nabla}_\perp\cdot {\bf v}_\perp$ to obtain by using the
kinematic boundary condition (\ref{kine})
\begin{eqnarray}
\frac{\partial h}{\partial t}&=&v_z(z=h)=-\frac{\xi\Delta\mu c_0
h_0^4}{8\eta}\nabla_\perp^4 h - \frac{(\lambda+\lambda_2)
h_0^3}{3\eta}\nabla_\perp^4 \psi - \frac{\xi\Delta\mu c_0
h_0^2}{6\eta} \nabla_\perp^2 h\nonumber \\&-&\frac{\sigma
h_0^3}{3\eta}\nabla^4_\perp h + \frac{\kappa h_0^3}{3\eta}
\nabla_\perp^6 h +\frac{\tilde\lambda \Delta\mu h_0^3}{3\eta}
\nabla^2_\perp \psi -\frac{\lambda_1 h_0^3}{3\eta} \nabla_\perp ^4
h, \label{hsol}
\end{eqnarray}
where we have linearised about the mean membrane height $h_0 =
\langle h\rangle$.
 Being a conserved density, the active density $\psi$ follows a
model B (in the nomenclature of Ref.~\cite{halp}) type equation,
together with advection. For an incompressible fluid and up to
linear order the equation takes the form (we have set a kinetic
coefficient to unity for simplicity and without any loss of
generality) \bea {\partial \psi \over
\partial t}= A \nabla_\perp^2\psi + E\nabla_\perp^4 h, \label{eqpsi}\eea where $E=(\lambda_2
+
\lambda)$. The dynamics of concentration $c$ follows the advection
equation
 \beq
 \frac{\partial c}{\partial t} + {\boldsymbol \nabla}\cdot [({\bf v}
 + v_0 {\bf p})c]=0,\label{eqc}
 \eeq
 where $v_0$ is a drift velocity. Writing $c=c_0 + c$, where  $c$ now refers to
 the (small) fluctuations of the local active particle concentration from
 $c_0$,
to the linear order in fluctuations (i.e., in ${\bf p}_\perp$ or
${\boldsymbol\nabla_\perp} h$ and $c$), $c$ follows the equation
 \beq
 \frac{\partial c}{\partial t}=-v_0 {\boldsymbol \nabla}\cdot ({\bf p}
 c)=c_0 v_0\nabla^2_\perp h,\label{eqc1}
 \eeq
where ${\bf p}_\perp$ has been replaced by its $z$-averaged
expression.
 Thus, the dynamics of $c$ is slaved to $h$-fluctuations.
Notice that at the linearised level and to the lowest order in
smallness, only $c_0$, the average active particle concentration in
the active fluid film, and not the fluctuations in $c$, enters into
the dynamics of $h$ and $\psi$. In contrast, the dynamics of $c$
itself is affected by fluctuations in $h$ and $\psi$. Hence, we
ignore the dynamics of $c$ and replace it by $c_0$ in the equations
for $h$ and $\psi$ for the remaining part of this article. Under the
$z$-averaged description used here, there is no independent dynamics
of ${\bf p}_\perp$. Thus, Eqs.~(\ref{hsol}) and (\ref{eqpsi}) for
$h$ and $\psi$ respectively together describe the effective dynamics
of the membrane-active fluid layer combine. This is in agreement
with our qualitative arguments above. Equations (\ref{hsol}) and
(\ref{eqpsi})  may be conveniently written in terms of spatially
Fourier-transformed variables, which are used below to elucidate the
linear instabilities:
\begin{eqnarray}
\frac{\partial h_{\bf q}}{\partial t}&=&\left[-\frac{\xi\Delta\mu
c_0 h_0^4}{8\eta} - \frac{(\sigma +\lambda_1)h_0^3}{3\eta} \right]
q^4 h_{\bf q} + \frac{\xi \Delta\mu c_0 h_0^2}{6\eta} q^2 h_{\bf q}
- \frac{\kappa h_0^3}{3\eta} q^6 h_{\bf q} -
\frac{\tilde\lambda\Delta\mu h_0^3}{3\eta} q^2 \psi_{\bf q} -
\frac{\lambda+\lambda_2}{3\eta} h_0^3 q^4 \psi_{\bf q},\label{hsolq}
\end{eqnarray}
and
\begin{equation}
\frac{\partial\psi_{\bf q}}{\partial t}= - Aq^2\psi_{\bf q} -
(\lambda +\lambda_2)q^4 h_{\bf q},\label{eqpsisolq}
\end{equation}
where $h_{\bf q}$ and $\psi_{\bf q}$ are spatial Fourier transforms
of $h({\bf r})$ and $\psi({\bf r})$; ${\bf q}=(q_x,q_y)$ is a $2d$
Fourier wavevector.

\subsection{Model II: The membrane active gel combine bounded by a
fluid interface below}

\subsubsection{Construction of the model and the free energy
functional}

\begin{figure}[htb]
 \includegraphics[height=6cm]{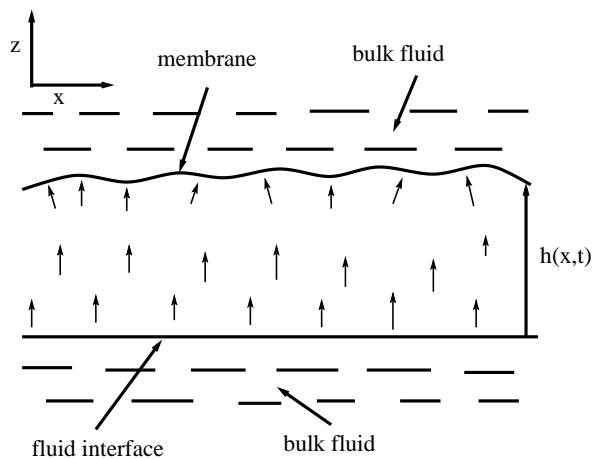}
 \caption{A schematic diagram of our model system showing the
membrane and the active fluid layer inside bulk passive fluid. The
arrows indicate the direction of macroscopic orientation (here along
the $z$-axis).}
 \label{liqfig}
 \end{figure}

Our calculation here broadly follows the framework outlined above. However, there are
 important
 differences in details, owing to symmetry considerations and boundary condition. We start by considering the free energy functional ${\cal F}_L$ of
the system. Assuming the system to be confined between $z=h_1$ and
$z=h_2$, ${\mathcal F}_L$ should be a functional of $h_1,\, h_2$,
$\bf p$ and $\psi$. This describes the energy of the system in a
given configuration defined by $\bf p$, $\psi, h_1$ and $h_2$. Just
as for ${\mathcal F}_s$, its form may be directly inferred from
symmetry considerations. It must generally be invariant under an
arbitrary tilt (equivalently a rotation) $h_{1,2} \rightarrow
h_{1,2} +{\bf a \cdot x}$ of the confining surfaces, where $\bf a$
is an arbitrary $3d$ vector and $\bf x$ is a $3d$ radius vector. At
this stage, similar to Ref.~\cite{niladri}, in order to simplify the
ensuing algebra, we assume that the interfacial tension of the lower
surface is large enough (formally diverging) so that its
fluctuations are strongly suppressed, and thus, $h_2$ drops out of
the dynamical description in this limit. Although this is not likely
to be directly realisable in any {\em in-vivo} or {\em in-vitro}
situations, they should nevertheless serve as a good starting point
for more refined and detailed theoretical modeling. To set our
notations simpler, we set $h_1 = h$ and $h_2 = 0$ below. See
Fig.~\ref{liqfig} below for a schematic representation of our model.

The tilt symmetry in Model II dictates that the most leading order
(in gradients) coupling bilinear in $\bf p$ and $h$  should be
through the local curvature. Hence, assuming small membrane
fluctuations, the most leading order term that contributes to the
relevant free energy functional ${\mathcal F}_L$ should be of the
form $\lambda_1{\boldsymbol\nabla}\cdot {\bf p}\nabla^2 h$. The
form of this
 $\lambda_1$-coupling term is a crucial difference with
${\mathcal F}_s$ [see (\ref{freesol}) above]; in order to avoid
introducing a large number of symbols, we continue to use the same
symbols as in (\ref{freesol}). Despite the structural differences
owing to symmetry considerations, the ${\bf p}-h$ and ${\bf p}-\psi$
couplings reflect the polarity of the system, similar to Model I. We
continue to represent the density of the active species on the
membrane by $\psi$. Similar to (\ref{freesol}), the free energy of
the active polar particles is given by the Frank free energy, which
we consider in the limit of equal Frank's constants, denoted by $D$
below, ignoring any anisotropy for simplicity. Thus the free
energy functional of the combined system of a fluid membrane, active
fluid layer and an active density on the membrane is given by
\begin{eqnarray}
{\cal F}_L (h,{\bf p},\psi)&=&{1 \over 2}\int d^2r [\sigma
({\boldsymbol\nabla}_\perp h)^2 + \kappa (\nabla_\bot^2h)^2  +
A(\psi -\psi_0)^2- 2\lambda\psi\nabla_\bot^2h\nonumber \\&+& {1
\over 2}\int d^2r\int_0^h dz [-\lambda_1 ({\boldsymbol\nabla}\cdot
{\bf p})\nabla^2_\perp h \delta(z-h) - 2\lambda_2 {\boldsymbol\cdot\bf p}
{\boldsymbol\nabla}_\bot \psi\delta(z-h) +D(\nabla_ip_j)^2],
\label{free}
\end{eqnarray}
where different symbols above have the same significance as in
(\ref{freesol}). We continue to impose $p^2=1$. Similar to Model I,
we have used the Monge gauge \cite{monge} above and have kept only
the terms which are either quadratic or bilinear in the fields.
  As
before, all the parameters in the model are chosen in such a way,
that a stable spatially uniform equilibrium phase~\cite{equl} (at
zero activity, $\Delta\mu=0$) is ensured. Our previous discussions
on the differences between ${\mathcal F}_s$ and the corresponding
free energy functional  in Ref.~\cite{niladri} similarly applies to
${\mathcal F}_L$ above and its corresponding free energy functional
in Ref.~\cite{niladri}.

\subsubsection{Boundary conditions}

Next, we specify the boundary conditions on the system: (i) At the
interfaces ($z = h$ and 0) polarisation ${\bf p}=(p_x,p_y,p_z)$ is
constrained to be perpendicular to the plane $z=0$ and the local
tangent plane on the surfaces $h=z$, i.e., ${\bf p}\cdot \hat N = 1$
at $z = h$, where $\hat N$ is the local normal at $h$ in the Monge
gauge (see above; equivalently at $z=h$ $p_i=-2\partial_i h,\,i=x,y$
for small membrane fluctuations and at $z=0, \,p_i=0$) and
$p_i=0,i=x,y$ at $z=0$; these are identical to the boundary
conditions used in Model I above, and (ii) continuity of the shear
stress at $z = h$ and 0, which for $\eta \gg \eta'$ reduces to the
condition (\ref{str}) at $z=h$ and $\eta\partial_z v_i =0$ at $z=0$,
where $i=x,y$. Notice the difference between the boundary
conditions on $v_i,i=x,y$ at $z=0$ for Model I and Model II: For
Model I, $v_i=0$ at $z=0$ (no-slip boundary condition), where as for
Model II, $\partial_z v_i =0$ at $z=0$. We shall see below that this
difference in the boundary conditions is responsible for the
differences in the fluctuation spectra that we obtain. The dynamics
of the height field $h$ is again determined by the kinematic
boundary condition~(\ref{kine}).

\subsubsection{The dynamical equations of motion}

We continue to use the active stress defined above (\ref{active})
that determines the active stress in the system. Similar to Model I
above, we assume no significant $z$-dependence of concentration $c$,
use a linear profile for ${\bf p}_\perp=-(2z{\boldsymbol
\nabla}_\perp h)/h$ that satisfies the boundary conditions imposed
on ${\bf p}_\perp$ and use this form to calculate $z$-averaged
active stresses from (\ref{active}). The validity of this approach
should still be the same as for Model I, the average thickness $h_0
=\langle h\rangle \ll \overline h$, the critical thickness at which
a spontaneous flow transition akin to the Frederiks transition of
equilibrium nematics.  For reasons similar to Model I, $h,\,\psi$
and $c$ will appear as the relevant dynamical fields in our
effective $2d$ description which we work out below.

As in the previous section, we use the generalised Stokes Eq. for
$3d$ velocity $v_\alpha, \alpha=x,y,z$. We further use the
Lubrication approximation for $v_z$ in the thin film limit yielding
an equation identical to Eq.~(\ref{pressS}).
 This can be integrated over $z$ to solve for
the pressure $\Pi$. The constant of integration is to be fixed by
the condition of normal stress balance. The pressure at the location
of the membrane is balanced by the total normal stress at $z=h$.
This includes the elastic force of the membrane and the active
stress due to the active density.  Thus, $\Pi$ is obtained as
 \bea
\Pi= P_0 -\xi c_0\Delta\mu(h-z)\nabla_\bot^2h - f(h) + \tilde\lambda\Delta\mu(\psi
-\psi_0), \label{pi} \eea where $f(h)$ is the elastic contribution
to the stress which can be derived from the free energy as
$f(h)=-\delta\mathcal{F}_L/\delta h=-\kappa \nabla_\bot^4h +
\sigma\nabla_\bot^2h - \lambda_1 \nabla_\bot^4 h +
(\lambda+\lambda_2)\nabla_\bot^2\psi$.

Since $v_i,i=x,y$ does not satisfy the no-slip boundary condition at
any of the confining surfaces at $z=0$ and $z=h$, unlike Model I,
$v_i$ cannot be obtained by using the lubrication approximation on
the corresponding Stokes' equation. Instead, we note that the zero
shear stress boundary conditions  at $z=0,h$ allows for a non-zero
$v_i$ that is {\em independent} of $z$, unlike Model I, which cannot
have a $z$-independent non-zero $v_i$. To proceed further and in the
spirit of a $z$-averaged description, we assume that $v_i$ has, in
addition to a $z$-independent part,  a $z$-dependent part with a
quadratic $z$-dependence $-\xi c_0\Delta\mu z^2\nabla_i h/(2h_0)$,
such that the boundary conditions on $\partial_z v_i$ are obviously
satisfied. Such an approach is physically meaningful over a
length-scale that is much larger than $h_0$, i.e., in terms of the
corresponding Fourier wavevector $\bf q$, $qh_0\ll 1$ (a more formal
derivation of the expression for $v_i$ is given in Appendix I). With
this we obtain in the Fourier space (a subscript refers variables in
the Fourier space)
\begin{eqnarray}
v_{i{\bf q}} &=& \frac{i\xi \Delta\mu c_0}{\eta q^2h_0} q_i h_{\bf q} -
\frac{iq_i}{\eta q^2}[\tilde\lambda\Delta\mu \psi_{\bf q} +\sigma
q^2 h_{\bf q} + \kappa q^4 h_{\bf q} + \lambda_1 q^4 h_{\bf q} \nonumber \\
&+&(\lambda +\lambda_2) q^2\psi_{\bf q} +\frac{\xi\Delta\mu c_0
h_0}{2} q^2 h_{\bf q}] + O(q^2 z^2).\label{vperp}
\end{eqnarray}
This, together with the incompressibility condition, then yields in
the Fourier space
\begin{eqnarray}
\frac{\partial h_{\bf q}}{\partial t}&=&{\xi\Delta\mu c_0 \over \eta} h_{\bf q}
-\frac{\tilde\lambda \Delta\mu h_0}{\eta}\psi_{\bf q} -
\frac{h_0}{\eta}\sigma q^2 h_{\bf q} - \frac{h_0}{\eta}
(\kappa+\lambda_1) q^4 h_{\bf q}\nonumber \\ && -
\frac{(\lambda+\lambda_2)h_0q^2}{\eta}\psi_{\bf q} - \frac{\xi\Delta\mu
c_0 h_0^2}{2\eta} q^2 h_{\bf q}.\label{height}
\end{eqnarray}
The dynamics of $\psi$ and $c$ follows the same equations
(\ref{eqpsi}) and (\ref{eqc1}) as for Model I. Again as in Model I,
the dynamics of $c$ is slaved to that of $h$-fluctuations in the
lowest order in smallness. Thus, equations (\ref{height}) and
(\ref{eqpsisolq}) together describe the effective $2d$ dynamics of Model
II in the long wavelength limit.

\section{Linear stability of polar ordered uniform states}
\label{instabil}

Having derived all the governing equations for the dynamics of Model
I and Model II, we perform linear stability analyses of small
fluctuations around the chosen ordered states.

\subsection{Stability analysis for Model I}

The linear stability of the chosen ordered state $p_z=1, h_{\bf
q}=0,{\bf p}_{\perp\, {\bf q}}=0,\psi_{\bf q}=0$ may easily be
ascertained by calculating the eigenvalues of the stability matrix
$M$ constructed from equations (\ref{hsolq}) and (\ref{eqpsisolq}):
 The eigenvalues are rather lengthy and are available in Appendix II.
 Regardless of the complicated structure of the eigenvalues $\Lambda$ as given in Eq.~(\ref{eigensolid}), we note that
 they vanish for ${\bf q}\rightarrow 0$, a consequence of screening of
the hydrodynamic interactions due to the presence of the solid
substrate below which generates friction. While straightforward
analysis of the eigenvalues (\ref{eigensolid}) requires considerable
algebraic manipulations, despite the complexity
 of the eigenvalues (\ref{eigensolid}) we can make the following
 observations: (i) There are no underdamped propagating modes, (ii)
 the system becomes unstable for either signature of $\Delta\mu$. The
 latter feature manifests itself clearly if we ignore the active density
 $\psi$ from the dynamics and analyse only the dynamics of $h$. We
 separately analyse for (i) $\psi=0$ and (ii) $\xi=0$.

\begin{figure}[htb]
\includegraphics[height=6cm]{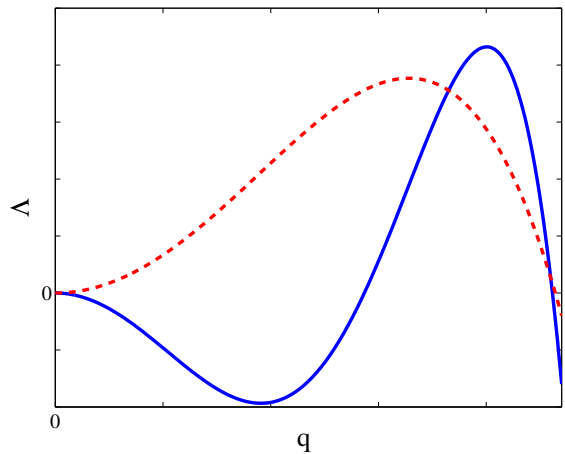}
\caption{Schematic plots of the eigenvalues $\Lambda$ of the linear
stability matrix as a function of wavevector $q$ in Model I when
$\psi=0$ for $\Delta\mu <0$ (blue continuous) and $\Delta\mu >0$
(red dashed line)}.\label{lambdasol}
\end{figure}

 Setting $\psi=0$, there is only one eigenvalue:
 \bea\Lambda={\xi c_0h_0^2q^2\Delta\mu \over
6\eta}-{\xi c_0\Delta\mu h_0^4q^4 \over 8\eta} - {(\sigma +\lambda_1)
h_0^3q^4 \over 3\eta}-\frac{\kappa h_0^3q^6}{3\eta}.\eea
 Thus for any $\Delta\mu >0$ (extensile active stress), the system is unstable
at $O(q^2)$, the lowest order in wavevector $\bf q$, but becomes
stable at $O(q^4)$. For a system with a linear lateral size $L$,
this yields an instability condition $L>L_c$, where the critical
linear size $L_c$ is given by
\begin{equation}
\xi\Delta\mu c_0 L_c^2 = \frac{3}{4}\xi\Delta\mu h_0^2 + 2 (\sigma
+\lambda_1)h_0.\label{ins1sol}
\end{equation}
Thus, for a given $\Delta\mu>0$ and $h_0$, there is always a system
size $L>L_c$, at which the instability will show up. We can
make an order of magnitude estimation for $L_c$ as follows: We take
$h_0\sim 100 nm$, a typical thickness of a cortical actin layer in a
cell, $\xi\Delta\mu\sim 7kCal/(500\times 10^{-23})$~\cite{niladri},
for an ordered system with a typical active particle size $a\sim
1nm$, $c_0a^3\sim 1$, $\sigma\tilde\lambda_1\sim 10mJ/m^2$. This
then yields a typical $L_c\sim 10^{-5} cm$, smaller than the linear
dimension of a cell. In contrast, for $\Delta\mu <0$ (contractile
active stress), there are no instabilities at $O(q^2)$, but the
system becomes unstable at $O(q^4)$ as soon as  as $h_0$ exceeds a
critical thickness $h_{0c}$ given by $\xi\Delta\mu_c c_0=
8(\sigma+\lambda_1)/(3h_0)$. Using the values of the parameters
as above we find, $h_{0c}\sim 10^{-6} cm$, smaller than the typical
thickness of a cortical actin layer. This, however, does not impose
any condition on system size $L$. Systems with a linear
dimension larger than  $L_c$ or with an average thickness larger
than $h_{0c}$ are expected to display the linear instabilities
obtained above. Of course, the system is stable at high enough $q$,
the bending modulus stabilises the system regardless of the sign of
$\Delta\mu$. The instabilities for either signature of $\Delta\mu$
may be understood as follows. In the underlying full $3d$ model,
there is only one nonequilibrium term, which is the active stress.
In the resulting $2d$ description, it contributes to the dynamics of
$h$ through (a) the $2d$ analogue of the active stress
(\ref{active}) and (b) the $2d$ active pressure. From the structure
of the generalised Stokes Eq. (\ref{stokesS}) for $v_z$, the $2d$
active pressure term balances the usual pressure, where as in the
Stokes Eq. for ${\bf v}_\perp$ the pressure and the $2d$ active
stress together balance the viscous stress term. Therefore, the
active pressure and the active stress terms appear with opposite
signs in ${\bf v}_\perp$ and hence in $v_z$ through the
incompressibility condition, and therefore in Eq.~(\ref{hsolq}) for
$h$. This leads to the instabilities for both signatures of
$\Delta\mu$ in the dynamics of $h$. A schematic diagram of the
eigenvalue $\Lambda$ when $\psi=0$ for both signs of $\Delta\mu$ is
given in Fig.~\ref{lambdasol}.

he instabilities are due to the $3d$ active stress term
$\xi\Delta\mu c_0 p_\alpha p_\beta,\,\alpha,\beta=x,y,z$, since the
second active term has been set to zero by setting $\psi=0$. The
opposite limit may also be examined by setting $\xi=0$, leaving the
active density-dependent term in the active stress expression
(\ref{active}) on the membrane as the only source of active stress
in the problem. One obtains coupled dynamical equations for $h$ and
$\psi$. The eigenvalues of the stability matrix are given by
\begin{eqnarray}
\Lambda&=&-A q^2 - \frac{\tilde\lambda \Delta\mu(\lambda+\lambda_2)
h_0^3}{3\eta A} q^4 +O(q^6); - (\sigma + \lambda_1)
h_0^3q^4+\frac{\tilde\lambda\Delta\mu (\lambda+\lambda_2)
h_0^3}{3\eta A} q^4 + O(q^6).
\end{eqnarray}
Thus, $\tilde\lambda\Delta\mu$ comes with opposite signs in the two
eigenvalues. Hence, for a fixed sign of $\tilde\lambda$ there are
instabilities associated with either signature of $\Delta\mu$.
However, the notable difference with $\psi=0$ is that the
instabilities now occur only at $O(q^4)$, much higher than $O(q^2)$.
In contrast, there is no instability $O(q^2)$, unlike the
instabilities which occur in the previous case ($\psi=0,\,\xi\neq
0$). In particular with positive $\tilde\lambda$, when $\Delta\mu
<0$, there is instability at $O(q^4)$ for any value of $|\Delta\mu
|$, where as when $\Delta\mu
>0$, there is instability at $O(q^4)$ provided $|\Delta\mu|$ exceeds
a critical value $\Delta\mu_c$ defined by $\tilde\lambda\Delta\mu_c
(\lambda +\lambda_2) = 3\eta A(\sigma+\lambda_1)$.  More generally,
in the present case, the cross-coupling between $h$ and $\psi$ is
responsible for the instabilities for both signatures of
$\Delta\mu$. While our detailed analysis of the eigenvalues above
rests on simplifying steps effectively involving keeping only one
term of the active stress expression (\ref{active}) at a time, we
expect the general conclusion on the presence of instability for
both signatures of $\Delta\mu$ should be valid. In particular, since
the governing equations of motion are linear, a combination of
 the instabilities elucidated above should be generally observed for proper choices of the model
 parameters when both $\xi$ and $\psi$
are non-zero. Finally, for high enough $q$, the system should always
be stabilised by the curvature contributions (not shown here).

\subsection{Stability analysis of Model II}
The instabilities to the uniform state $h_{\bf q}=0,\,{\bf
p}_{\perp\,{\bf q}}=0,\psi_{\bf q}=0$ may be obtained by calculating
the eigenvalues of the stability matrix from the equations
(\ref{eqpsisolq}) and (\ref{height}). The full expressions of the
eigenvalues are given in Appendix~\ref{appen3}, which are rather
lengthy. However, we use their forms up to $O(q^2)$ for our analyses
in this Section below. These instabilities have both qualitative
similarities and dissimilarities with those in Model I which are
discussed in details below. As before, we analyse the eigenvalues
for two special cases - (i) $\psi=0$, and (ii) $\xi=0$.

\begin{figure}[htb]
\includegraphics[height=6cm]{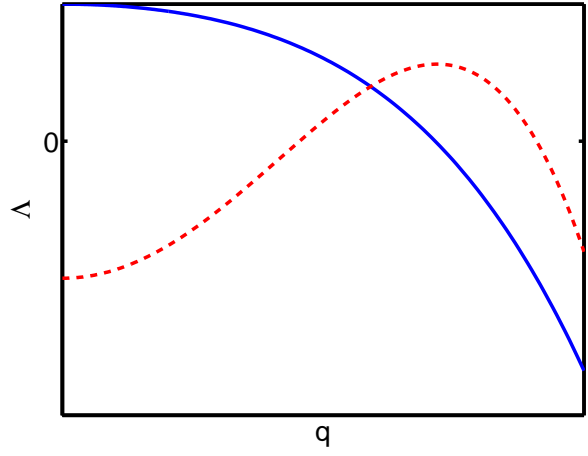}
\caption{Schematic plots of the eigenvalues of the linear stability matrix in Model II
as a function of wavevector $q$ when $\psi=0$ for $\Delta\mu >0$
(blue continuous) and $\Delta\mu <0$ (red dashed
line)}.\label{lambda}
\end{figure}

With $\psi=0$, the height field $h$ remains the only relevant field
and we obtain
\begin{eqnarray}
\Lambda={\xi\Delta\mu c_0 \over \eta}- \frac{h_0\sigma q^2}{\eta} -
\frac{h_0(\kappa+\lambda_1)}{\eta} q^4 h -\frac{\xi\Delta\mu c_0
h_0^2}{2\eta} q^2, \label{eigenliqh}
\end{eqnarray}
Thus, with $\Delta\mu >0$ (extensile), the system is unstable at
$O(q^0)$, but stable at $O(q^2)$ or higher. Similar to Model I, the
instability condition (\ref{eigenliqh}) imposes a condition on the
system size $L$ for instability: One must have
\begin{equation}
L^2 > [h_0\sigma + \xi\Delta\mu c_0h_0^2/2]/(\xi\Delta\mu c_0)
\label{ins1liq}
\end{equation}
for instability. Condition (\ref{ins1liq}) is analogous to the
condition (\ref{ins1sol}) for Model I above.
 In contrast, with $\Delta\mu <0$, (contractile)
the system is stable at $O(q^0)$, but becomes unstable at $O(q^2)$
if $|\xi\Delta\mu c_0|> \frac{2}{h_0}\sigma $. As for Model I, there
is no condition on the system size $L$ for occurrence of the
instability.  Thus the system may display instabilities for both
signatures of $\Delta\mu$, again very similar to the corresponding
result from Model I. The physical origin of these instabilities are
again same as that for Model I, {\em viz}, that the $3d$ active
stress (\ref{active}), with $\psi=0$, contributes to the $2d$
analogue of (\ref{active}) and also makes an {\em active}
contribution to the pressure (\ref{pi}). For high enough $q$, the
instabilities are always suppressed by the curvature contribution,
regardless of the signature of $\Delta\mu$.
 A schematic diagram of the
eigenvalue $\Lambda$ when $\psi=0$ for both signs of $\Delta\mu$ is
given in Fig.~\ref{lambda}.

The other limiting case with $\xi=0$, but $\psi\neq 0$ turns out to
be qualitatively different from its counterpart in Model I. There
are two eigenvalues, for there are now two dynamical fields, $h$ and
$\psi$. The eigenvalues obtained are available in Appendix III; to
the lowest order in $O(q^2)$ they are
\begin{eqnarray}
\Lambda&=& -\frac{1}{2}[\frac{h_0}{\eta}\sigma q^2 + A q^2] \pm
\frac{q^2}{2}[(A-\frac{h_0}{\eta}\sigma )^2 + 4{(\lambda
+\lambda_2)\tilde\lambda\Delta \mu h_0 \over \eta}]^{1/2} + O(q^4).
\label{eigenliq2}
\end{eqnarray}
From (\ref{eigenliq2}) note that various scenarios are possible,
which we consider by taking various limits (we assume
$A>h_0\sigma/\eta$ for simplicity). Two different possibilities
exist: (a) When $(A-\frac{h_0\sigma}{\eta})^2 \gg 4 {(\lambda
+\lambda_2)\tilde\lambda\Delta\mu h_0 \over \eta}$ then,
\begin{itemize}
\item If $\frac{(\lambda +\lambda_2) \tilde\lambda\Delta\mu}{A-\frac{h_0\sigma}
{\eta}} > \sigma$, there is an instability at $O(q^2)$ with
positive $\Delta\mu$.
\item If $\frac{(\lambda +\lambda_2)|\tilde\lambda \Delta\mu
|}{A-\frac{h_0\sigma}{\eta}} > A{\eta \over h_0}$, the system is unstable for
$\Delta\mu <0$ with an instability at $O(q^2)$.
\end{itemize}
There are no underdamped propagating modes in the system. In
contrast (b) when $4{(\lambda + \lambda_2)\tilde\lambda\Delta\mu h_0
\over \eta} \gg (A-\frac{h_0\sigma}{\eta})^2$ then
\begin{itemize}
\item There are underdamped propagating modes for any $\tilde \lambda\Delta\mu
<0$ with dispersion proportional to $q^2$ and speed depending
linearly on $\Delta\mu$. In this case, however, there are no
instabilities.
\item When $\tilde\lambda\Delta\mu >0$, there are instabilities at $O(q^2)$
when $[(\lambda +\lambda_2)\tilde\lambda\Delta\mu]^{1/2} +
\frac{(A-h_0\sigma/\eta)^2}{8\sqrt{(\lambda+\lambda_2)\tilde\lambda\Delta\mu}}>
[\frac{h_0\sigma}{\eta} +A]/2$.
\end{itemize}
 Similar to Model I, in a general situation with both $\xi$ and
$\psi$ being non-zero, a combination of all the linear instabilities
discussed above should be displayed by the system with appropriate
choices for the model parameters.


Let us now look at the similarities and differences between the mode
spectra and instabilities in Model I and Model II. When the active
density $\psi$ is absent, both Model I and II yield generic
instabilities for both signatures of $\Delta\mu$ for the dynamics of
$h_{\bf q}$. There are, however, significant differences too. At a
technical level, first of all, in Model II, the eigenvalues {\em do
not vanish} for $ q\rightarrow 0$. This is a consequence of the
Stokesian dynamics for the velocity field; see Ref.~\cite{niladri}
for more discussions in an analogous problem. This is in contrast to
Model I, where hydrodynamic interactions are screened and the
eigenvalues smoothly go over to zero as $q\rightarrow 0$. There are
other differences which manifest when $\psi\neq 0$, leading to the
existence of underdamped propagating waves in Model II, unlike Model
I where there is no such propagating mode. This is formally  a
consequence of the Stokesian dynamics together with hydrodynamic
interactions in Model II: The active stress contribution from the
density $\psi$ to the  dynamics od $h$ for Model II yields {\em
leading order} contribution (along with other leading order terms),
where as for Model I, such contributions are always subleading. This
difference leads to the existence of underdamped propagating modes
for $\Delta\mu <0$ in Model II, a possibility that does not exist in
Model I.



\section{Summary and outlook}
\label{conclu}

In summary, thus, we have formulated an effective $2d$ description
for the coupled dynamics of small fluctuations of height $h$ and
active species density $\psi$ in a thin layer of fluid
membrane-active fluid combine about a reference state with $p_z=1$,
$z$ being the thin direction. Linear stability analysis of the model
reveals possibilities for instabilities for either signature of
$\Delta\mu$. These instabilities may be moving, as in our Model I
with a solid substrate below, or static (or localised) as in our
Model II with the membrane-active fluid combine being embedded in a
bulk fluid. We have analysed the roles played by the two sources of
nonequilibrium active stresses separately in generating the linear
instabilities in our models. Our results amply highlight the role of
boundary conditions in determining the long wavelength properties of
the system.

Direct comparisons of our results  with the available experimental
results are difficult, mainly due to the simplified, minimalist
nature of our model and (to our knowledge) the lack of precise
measurements of membrane fluctuations. Nevertheless,
Ref.~\cite{koend} reports an unusually low value for the surface
tension obtained for a cytoskeletal actin-myosin network
encapsulated in giant liposomes by fitting the membrane fluctuations
with the Helfrich model, which does not take active effects into
account. While the presence of active stresses and the consequent
instabilities may well be responsible in such an unusual value for
the (effective) surface tension, we cannot say anything conclusively
at present. Further experimental works may be necessary for
resolving this issue definitively. However, regardless of the
present status of the experimental results, {\em in-vitro} systems
such as those used in Refs.~\cite{bass,koend} may in principle be
used to study our theoretical predictions. Our estimation of
$L_c\sim 10^{-5}cm$ and $h_{0c}\sim 10^{-6} cm$, while indicating
that a system size smaller than a typical eukaryotic cell and/or
thinner than a cortical actin layer should be linearly unstable and
will not stay in a linearly ordered state, are only suggestive and
any quantitative accuracy is not expected. Since our calculational
framework is  applicable only when the average system thickness is
much smaller than the threshold of the spontaneous flow transition,
we cannot comment on the linear stability of system near the
threshold. We note here that the results from our Model II are
largely of theoretical interests. This is primarily because of our
assumption of a diverging interfacial tension at the bottom surface
of the system. In a real situation when the interfacial tension is
generically finite,  the bottom surface  fluctuates, and one
generally has fluctuations of both the mean height $(h_1+h_2)/2$ and
thickness $h_1-h_2$.  Our model does not capture such features. Full
$3d$ calculations are needed to capture these additional features.
Nevertheless,  qualitative aspects of our results should be visible
in more refined calculations.

Our results are complementary to those from related theoretical
works. For instance, Ref.~\cite{nir} in a coarse-grained linearised
model in terms of the membrane height field $h$ and the active
protein density $n$ ($\psi$ in our notation) showed, along with
linear instabilities, how the membrane gets a mean velocity in the
model: $\langle \frac{\partial h}{\partial t}\rangle$ is non-zero
and taken to be proportional to $n$ in their model, which is
physically due to the protruding forces coming from the
polymerisation of the normally anchored actin filaments. While we do
not have  any overall movement of the membrane in our model, this
can be included easily in our model by an explicit addition of a
term proportional to $\psi$ in Eqs.~(\ref{height}) and (\ref{hsol}).
However, this should be done with caution: If there is a constant
(say, upward according to our Fig.~\ref{liqfig}) velocity of the
membrane, the top-to-bottom distance will rise indefinitely and our
$z$-averaged effective $2d$ description will become invalid
eventually. In a related $2d$ model, Ref.~\cite{nir1} extends the
model of Ref.~\cite{nir} by including the effects of contractile
forces due to molecular motors such as myosin, leading to generic
travelling waves in the membrane. In contrast, our model (Model I)
displays no underdamped travelling waves. Unlike
Refs.~\cite{nir,nir1}, our model explicitly incorporates orientation
fluctuation-dependent active stresses and displays instability for
both contractile and extensile active stresses (both signs of
$\Delta\mu$).  Lastly, instead of a one-component active polar
species one may consider {\em active protein pumps}, modelled by
densities of upward and downward pumps, together with
permeation~\cite{sriram-pump}. These are known to introduce
travelling waves or instabilities in the membrane, depending on
details (e.g., local structural or functional asymmetries).  We hope
our work will induce further theoretical and experimental studies
along these directions.

\section{Appendix I: Alternative derivation of ${\bf v}_{i\bf q}$ in
Model II}\label{appen1}

The in-plane velocity $v_{i\bf q}$ for Model II may be obtained
using the Green's function technique. Let  $G_q(z,z')$ Green's
function for Eq.~(\ref{stokesS})
 such that it satisfies an equation \bea \eta {\partial^2G_q(z,z') \over
\partial z^2} - \eta q^2G_q(z,z')= \delta(z-z'). \label{greenstokes}
\eea Further, $G_q$ obeys the same boundary conditions as $v_{i\bf
q}$ for Model II. Then for $z\neq z'$, the solutions for $G_q(z,z')$
are given by \bea
G_q(z>z') &=& A_1\sinh{qz}+B_1\cosh{qz} \\
G_q(z<z') &=& A_2\sinh{qz}+B_2\cosh{qz}. \eea From the boundary
conditions on $v_{i\bf q}$, we obtain $\partial_zG_q(z<z')=0$ at
$z=0$ and $\partial_zG_q(z>z')=a(x,y)$ at $z=h$, where $a$ is any
function of $x$ and $y$, which is to be determined. Thus we obtain
\bea
G_q(z<z') &=& B_2\cosh{qz}, \\
G_q(z>z') &=& {a\sinh{qz} \over q\cosh{qh}} + {B_1\cosh{(z-h)} \over \cosh{qh}}.
\eea

Integrating Eq.~(\ref{greenstokes}) we get
\bea
\eta\left({\partial G_q \over \partial z}\right)_{z=z'+\epsilon} - \eta\left({\partial G_q \over
\partial z}\right)_{z=z'-\epsilon} =1.
\eea
This along with the continuity of the Green's function solution i.e.,
$G_q|_{z=z'+\epsilon}=G_q|_{z=z'-\epsilon}$ gives us
\bea
G_q(z>z') &=& -{\cosh{qz'}\cosh{q(z-h)} \over \eta q\sinh{qh}} + {a\cosh{qz}
\over q\sinh{qh}}, \\
G_q(z<z') &=& -{\cosh{q(z'-h)}\cosh{qz} \over \eta q\sinh{qh}} + {a\cosh{qz}
\over q\sinh{qh}}.
\eea

Now we can write $v_{i\bf q}$ as \bea
v_{i\bf q} &=& \int_0^h G_q(z,z')\phi_{i\bf q}(z')dz' \\
&=&\int_0^z G(z<z')\phi_{\bot q}(z') dz' + \int_z^h
G_q(z>z')\phi_{i\bf q} (z') dz' \eea where the kernel (in the
Fourier space) $\phi_{i\bf q}=iq_i\Pi+iq_j\sigma_{i j}$ can be
evaluated using Eqs.~(\ref{active}) and (\ref{pi}). Eliminating
$a(x,y)$ using $\eta\partial_zv_{i\bf q}|_{z=h}=-i\xi\Delta\mu
c_0h_0 qh$, the final expression for $v_{1\bf q}$ comes out to be
(in the limit $qh \ll 1$ and retaining the lowest order terms
  \bea v_{i{\bf q}} &=& \frac{i\xi \Delta\mu c_0}{\eta q^2h_0} q_i
h_{\bf q} - \frac{iq_i}{\eta q^2}[\tilde\lambda\Delta\mu \psi_{\bf
q} +\sigma
q^2 h_{\bf q} + \kappa q^4 h + \lambda_1 q^4 h_{\bf q} \nonumber \\
&+&(\lambda +\lambda_2) q^2\psi_{\bf q} +\frac{\xi\Delta\mu c_0
h_0}{2} q^2 h_{\bf q}] + O(q^2 z^2). \eea upon linearisation, where
${\bf q}$ is a two dimensional fourier wavevector. This is same as
Eq.~(\ref{vperp}).

\section{Appendix II: Eigenvalues of the stability matrix in
Model I}\label{appen2}

The eigenvalues are formally given by
 \bea \Lambda
&=&\left[M_{11}+M_{22} \pm \{(M_{11}+M_{22})^2 - 4 (M_{11}M_{22} - M_{12}M_{21})\}^{1/2}\right] , \label{eigensolid}\nonumber \\
\eea
 where $$M_{11}=\frac{\xi\Delta\mu c_0 h_0^2}{6\eta} q^2 -
 \left[\frac{\xi\Delta\mu c_0 h_0^4}{8\eta} +\frac{(\sigma+\lambda_1) h_0^3}{3\eta}
\right]  -\frac{\kappa h_0^3}{3\eta}
 q^6,$$
 $$M_{12}=-\frac{\tilde\lambda\Delta\mu h_0^3}{3\eta}q^2 -
 \frac{(\lambda+\lambda_2) h_0^3}{3\eta} q^4,$$
 $$M_{21}=-(\lambda + \lambda_2) q^4,$$
 $$M_{22}= -A q^2,$$
where $M_{11},M_{12},M_{21},M_{22}$ are the different elements of
the $2\times 2$ stability matrix $M$.

\section{Appendix III: Eigenvalues of the stability matrix in
Model II}\label{appen3}

The eigenvalues of the stability matrix for $\xi=0$ and $\psi\neq 0$
in Model II can be written explicitly as
\bea
\Lambda &=& -{1 \over 2}\left[\left(A+{\sigma h_0 \over \eta}\right)q^2+
{(\kappa +\lambda_1)h_0q^4 \over \eta}\right] \nonumber \\
&&\pm \left[\left(A-{\sigma h_0
\over \eta}\right)^2q^4\left\{1-{(\kappa +\lambda_1)h_0q^2 \over \eta(A-{\sigma h_0
\over \eta})}\right\}^2+4(\lambda +\lambda_2){\tilde\lambda\Delta\mu h_0q^4
\over \eta} + 4(\lambda +\lambda_2)^2{h_0q^6 \over \eta}\right]^{1 \over 2}.
\eea
Expanding the above to $O(q^2)$ yields Eq.~(\ref{eigenliq2}).

\section{Acknowledgement}
AB gratefully acknowledges partial financial support in the form of
the Max-Planck Partner Group at the Saha Institute of Nuclear
Physics, Calcutta, funded jointly by the Max-Planck- Gesellschaft
(Germany) and the Department of Science and Technology (India)
through the Partner Group programme (2009).

\end{document}